\documentstyle[11pt,aaspp4]{article}
\received{}
\accepted{}
\input tex.def
\slugcomment{To appear in {\it The Astronomical Journal}.}

\begin{document}

\title{Discovery of Radio Outbursts in the Active Nucleus of M81}

\author{Luis C. Ho}
\affil{Carnegie Observatories, 813 Santa Barbara St., Pasadena, CA 91101-1292}

\author{Schuyler D. Van Dyk}
\affil{Infrared Processing and Analysis Center, California Institute of 
Technology, Mail Code 100-22, Pasadena, CA 91125}

\author{Guy G. Pooley}
\affil{Mullard Radio Astronomy Observatory, Cavendish Laboratory, Madingley
Road, Cambridge, CB3 0HE, England}
 
\author{Richard A. Sramek}
\affil{National Radio Astronomy Observatory, P.O. Box 0, Socorro, NM 87801}

\and 

\author{Kurt W. Weiler}
\affil{Remote Sensing Division, Naval Research Laboratory,
Code 7214, Washington, DC 20375-5320}

\begin{abstract}
The low-luminosity active galactic nucleus of M81 has been monitored at 
centimeter wavelengths since early 1993 as a by-product of radio programs to 
study the radio emission from Supernova 1993J.  The extensive data sets reveal 
that the nucleus experienced several radio outbursts during the monitoring 
period.  At 2 and 3.6~cm, the main outburst occurred roughly in the beginning 
of 1993 September and lasted for approximately three months; at longer 
wavelengths, the maximum flux density decreases, and the onset of the burst is 
delayed.  These characteristics qualitatively resemble the standard model for 
adiabatically expanding radio sources, although certain discrepancies between 
the observations and the theoretical predictions suggest that the model is 
too simplistic.  In addition to the large-amplitude, prolonged variations, we 
also detected milder changes in the flux density at 3.6~cm and possibly at 
6~cm on short (\lax 1 day) timescales.  We discuss a possible association 
between the radio activity and an optical flare observed during the period 
that the nucleus was monitored at radio wavelengths.
\end{abstract}


\section{Introduction}

At a distance of only 3.6 Mpc (Freedman \etal 1994), NGC 3031 (M81) hosts 
perhaps the best studied low-luminosity active galactic nucleus (AGN).  Its 
striking resemblance to ``classical'' Seyfert 1 nuclei was first noticed 
by Peimbert \& Torres-Peimbert (1981) and Shuder \& Osterbrock (1981), and 
a number of subsequent studies (Filippenko \& Sargent 1988; 
Keel 1989; Ho, Filippenko, \& Sargent 1996) have elaborated on its AGN-like 
characteristics.  The nucleus of M81 holds additional significance because 
of its relevance to a poorly understood class of emission-line objects 
known as low-ionization nuclear emission-line regions (LINERs: Heckman 1980), 
whose physical origin is still controversial (Ho 1999 and 
references therein).  The optical classification of the nucleus of M81 borders 
between LINERs and Seyferts (Ho, Filippenko, \& Sargent 1997), but it is clear 
that the ionization state of its line-emitting regions differs significantly 
from that of typical Seyfert nuclei (Ho \etal 1996).

The variability properties of low-luminosity AGNs, LINERs in particular, are 
very poorly known at any wavelength.  The faintness of these nuclei renders 
most observations extremely challenging, and routine monitoring of them has 
rarely been attempted.  M81 remains one of the few LINERs 
with sufficient data to allow its variability characteristics to be assessed.  
The occurrence of Supernova (SN) 1993J in late March of 1993 prompted us almost 
immediately to monitor the supernova interferometrically
at radio wavelengths.  Since at most wavelengths the field of view of the 
telescopes contained both the supernova and the center of the galaxy, the 
observations yielded a useful record of the radio flux density of the nucleus 
during the period that the supernova was monitored.  This paper will 
examine the radio variability properties of the nucleus of M81 based on these 
data.  We report on the discovery of several radio outbursts, which 
plausibly could be associated with the optical flare seen by Bower 
\etal (1996).  

\section{Radio Observations}

The radio data analyzed in this paper are based on observations of 
SN 1993J made with the Very Large Array (VLA)\footnote{The VLA is a facility 
of the National Radio Astronomy Observatory which is operated by Associated 
Universities, Inc., under cooperative agreement with the National Science 
Foundation.} as reported by Van Dyk \etal (1994), on similar monitoring data 
acquired using the Ryle Telescope by Pooley \& Green (1993), and on 
unpublished updates to these two data sets since then.  

\subsection{VLA Data}

The VLA data were acquired in snapshot mode at 20~cm (1.4 GHz), 6~cm 
(4.9 GHz), 3.6~cm (8.4 GHz), 2~cm (14.9 GHz), and 1.3~cm (22.5 GHz), using a 
50 MHz bandwidth for each of the two IFs, between 1993 March 31 and 1997 
January 23.  The different bands were observed nearly simultaneously.  As 
scheduling constraints did not permit us to specify in advance the array 
configuration for the observations, the data were taken using a mixture of 
configurations and hence a range of angular resolutions (synthesized beam 
0\farcs25--44\asec).  Because of the coarseness of the beam in the most 
compact configurations and at the longest wavelength, extended, off-nuclear 
emission can potentially contaminate the signal from the central point source. 
Inspection of the detailed maps of Kaufman \etal (1996), however, indicates 
any such contamination of the nucleus, even at 20~cm, is at most a few 
percent.  Since the phase center of the VLA observations was 
near the position of SN 1993J, and the nucleus of M81 is 2\farcm64 from this 
center, primary beam attenuation had a critical effect on the observations, 
such that only the 20, 6, and 3.6~cm observations produced reliable 
data for the nucleus.  At 2~cm the nucleus was too close to the edge of the 
primary beam to produce reliable measurements for all configurations.
At 1.3~cm the nucleus was outside the primary beam, and therefore maps were 
not made of the nucleus at this wavelength.

VLA phase and flux density calibration and data reduction followed 
standard procedures within the Astronomical Image Processing
System (AIPS), such as those described by Weiler et al. (1986),
using 3C~286 as the primary flux density calibrator and 1044$+$719 as the 
main secondary flux density and phase calibrator.  For observations at 
20~cm in the more compact D configuration, 0945$+$664 was the secondary 
calibrator.  The flux density scale is believed to be consistent with that 
of Baars \etal (1977).

Maps were made of the 20, 6, and 3.6~cm observations in the usual manner 
within AIPS, offsetting the map center from the observation phase-tracking 
center to the position of the nucleus.  Since the supernova near maximum was 
nearly as bright as the nucleus ($\sim$100 mJy), the sidelobes of  SN 1993J 
can severely contaminate the measurements of the nucleus.  The size of the 
map for each frequency and configuration was chosen so that both sources 
were included in the field, and the map was then deconvolved using the 
CLEAN algorithm (as implemented in the task IMAGR).  We determined the 
depth of the cleaning by empirically examining the convergence of the 
recovered flux density.  Both the peak and the integrated flux densities of 
the nucleus on the deconvolved maps were measured by putting a tight box around 
the source and summing the pixel values using the task IMEAN.  Because 
the nucleus is displaced so far from the phase center, the image of the 
nucleus is affected at all frequencies by bandwidth smearing, which diminishes 
the peak flux density relative to the integrated flux density, a conserved 
quantity.  Therefore, we report here only the integrated flux densities.  We 
verified that the more sophisticated procedure of fitting the source with 
elliptical Gaussians (using the task IMFIT) gives essentially the same 
results (usually within 3\%).  The final flux densities were obtained by 
scaling the observed values with the correction factor for the attenuation 
of the primary beam.  For the 25~m dishes of the VLA, the correction factors at 
20, 6, and 3.6~cm are 1.008, 1.242, and 2.011, respectively, for a 
phase-center displacement of 2\farcm64.

Because we are interested in establishing the variability behavior of the 
nucleus using a data set taken under nonoptimal conditions, it is vital 
to make a careful assessment of the various sources of uncertainty that 
can affect the measurements.  The following sources of error are considered:

(1) The quality of the maps varies widely, but in general the rms noise in the 
vicinity of the nucleus ranges from 0.1 to 2 mJy beam$^{-1}$, with median 
values of 0.7, 0.3, and 0.4 mJy at 20, 6, and 3.6~cm, respectively.  The 
corresponding fractional error on the total flux densities is 0.8\%, 0.4\%, and 
0.6\% at 20, 6, and 3.6~cm, respectively.  

(2) As mentioned above, the process of measuring flux densities from the maps 
itself can introduce an uncertainty as large as $\sim$3\%, depending on the 
method adopted (IMFIT or IMEAN).  

(3) The absolute flux density scale of the primary calibrator is assumed 
accurate to better than 5\% (see, e.g., Weiler \etal 1986).  

(4) The primary beam of the VLA antennas is accurate only to a few percent at 
the half power point (Napier \& Rots 1982); the exact uncertainty, in fact, 
has not been determined rigorously (R. Perley and M. Rupen 1999, private 
communications).  For concreteness, we will assume that this source of 
uncertainty contributes an error of 5\% to the primary-beam correction at all 
frequencies.  

(5) Without reference pointing, the VLA can achieve rms pointing errors, under 
good weather conditions, of 15\asec--18\asec\ during the night and 
\gax20\asec\ during the day (Morris 1991).  At the half power point of the 
beam, a pointing error of 20\asec\ results in an amplitude error of 
18.4\% at 3.6~cm after accounting for primary beam corrections.  Assuming that
this error is independent among the 27 antennas, we can divide it by $\sqrt{27}$
to get a 3.5\% error in the flux measurement of a source at the beam half power 
point due to pointing.  A similar calculation yields an error of 1.3\% at 
6~cm and 0.09\% at 20~cm.

(6) A potentially more serious source of uncertainty comes from systematic 
pointing errors induced by wind and solar heating; these errors are not 
expected to be random among the antennas.  According to Morris 
(1991), a wind speed of 8 m~s$^{-1}$ (a typical value for windy conditions) 
introduces an additional pointing error of $\sim$20\asec.  The formal error 
due to differential solar heating is difficult to establish because this 
effect has not been formally quantified for the VLA, but it has been estimated 
to be a factor of a few smaller than the contribution from moderate winds.  We 
take 20\asec\ as a nominal pointing error for the two contributions.  After 
accounting for primary beam correction, this translates into a flux 
measurement error of 18\% at 3.6~cm, 7\% at 6~cm, and 0.5\% at 20~cm.  

(7) The problem of ``beam squint'' --- the variation of the primary beam 
caused by slight differences between the pointing centers of the right and left 
polarizations --- is negligible ($<$0.5\%) because we make use only of the 
total intensity, the sum of both polarizations.  We neglect this 
contribution to the total error.

A reasonable estimate of the final error budget can be derived by summing in 
quadrature sources (1) through (5).  This corresponds to an uncertainty 
of approximately 8\% for all three frequencies.
%
%8.5\% at X; 7.8\% at C 7.7\% at L
%
The true uncertainty of any individual measurement at 3.6~cm, on the 
other hand, can be substantially larger than this if systematic pointing 
errors induced by wind or solar heating are significant.  In the most extreme 
situation, the total error could be as large as 20\%.   

\subsection{Ryle Telescope Data}

The Ryle Telescope (the upgraded Cambridge 5-km Telescope; Jones 1991) is an 
E-W synthesis telescope at the time of the observations operating at 15.2 GHz 
with a bandwidth of 280 MHz. The angular separation between the supernova and the nucleus of M81 
places one near the half-power point when the pointing center is at the other. 
The disk of the galaxy also gives rise to emission on similar angular scales.
In order to make a clean distinction between the two responses, and
to reject the emission from the disc,  we cannot make use of the 
short baselines ($<$100 m) in the array: the resolution is inadequate. 
We therefore use the two groups of longer baselines, near 1.2 and 2.4 km, 
with interferometer fringe spacings of 2\farcs6 and 1\farcs3.  So long as 
we avoid the hour angles at which the fringe rates were similar, integration 
removes the response to any source which is not at the phase center.
For the first month after the supernova explosion no observations
centered on the nucleus were made; we have not attempted to analyze the data
during this interval to derive a flux density for the nucleus. From 1993 May 
5 until 1994 June 20, separate pointings were made on the supernova and on 
the nucleus, together with one on the nearby quasar B0954+658 as a phase 
calibrator.  Amplitude calibration is based on  observations of 3C~48 or 
3C~286, normally on a daily basis.  The data presented here were made with 
linearly-polarized feeds and are measurements of Stokes I+Q.  We estimate that 
the typical rms uncertainty in the flux densities is $\sim$5\%.  
%It is clear from the day-to-day calibration observations that there were some 
%residual fluctuations in the flux-density scale. 
It is apparent from the observations of the phase calibrator, the supernova
and the nucleus that some fluctuations in the amplitude scale remain; these
are a consequence of poor weather conditions during which the system noise
and the gain of the telescope are subject to variations which have not been
completely removed by the monitoring systems.  As discussed by Pooley \& Green 
(1993), the quasar B0954+658 is strongly variable (a factor of 2 at 15 GHz 
over the interval covered here), and we fitted a smoothly-varying curve to 
its apparent flux density in order to reduce the effects of weather and 
system fluctuations on the results. In retrospect, using a calibrator
whose flux density was varying less dramatically would have been better.

\section{The Light Curves}

The light curves of the nucleus (Fig. 1) exhibit a complex pattern of flux 
density variations during the monitoring period.  The nucleus emits a baseline 
level of roughly 80--100 mJy at all four wavelengths, consistent with its 
historical average (Crane, Giuffrida, \& Carlson 1976; de~Bruyn \etal 1976; 
Bietenholz \etal 1996), on which is superposed at least one, and possibly 
three to four, discrete events during which the flux increased substantially.  

\begin{figure}
%\plotone{figs/fig1_v8_portrait.ps}
\plotone{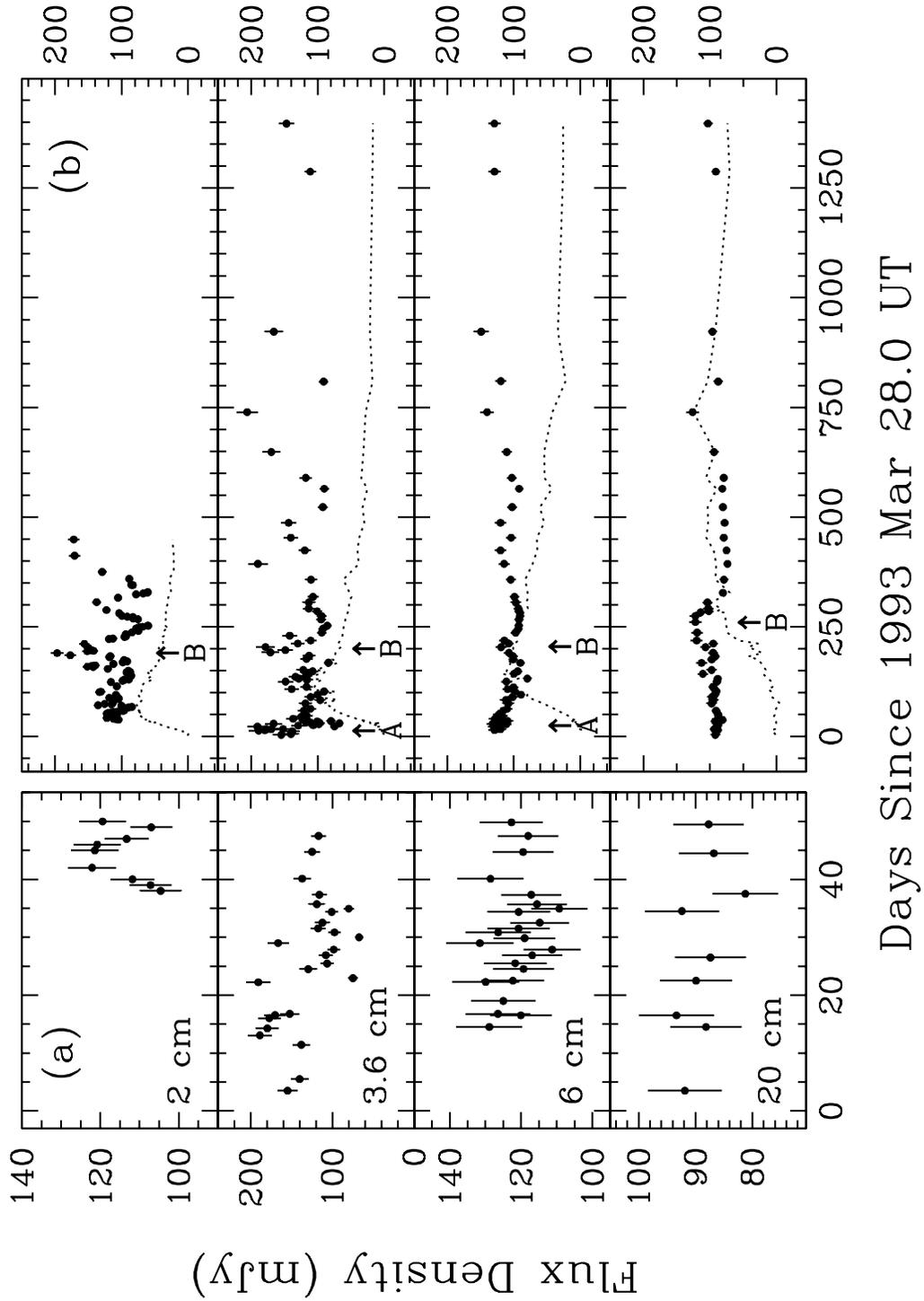}
\caption{
Radio light curves of the nucleus of M81.  The abscissa denotes days since
the explosion of SN 1993J, which we adopt as 1993 March 28.0 UT.  The flux
densities for the first 55 days ({\it a}) are shown separately from the full
data set ({\it b}) in order to highlight the short timescale variations
discernible in the closely-sampled part of the light curve.  The dashed
lines represent the light curves of SN 1993J measured from the same images.
Outbursts A and B are marked with arrows.
}
\end{figure}

The ``outbursts'' are most clearly visible in the VLA data at 3.6~cm.  
Although our earliest epoch did not catch the onset of the initial rise of the 
first event, a local maximum is apparent on day\footnote{The light curves will 
be referenced to days since 1993 March 28.0 UT, the adopted date of the 
explosion of SN 1993J.} 13$\pm$5 (hereafter outburst ``A;'' Fig. 1{\it a}).  
This outburst reached a peak flux density of 190 mJy, a nearly two-fold 
increase in brightness over the quiescent state.  There is some indication 
that outburst ``A'' is present at 6~cm as well; however, it appears to be 
slightly delayed with respect to 3.6~cm, and the amplitude of the intensity 
increase compared to the quiescent level is lower, on the order of 40\%.   We 
find no convincing evidence for a corresponding variation at 20~cm, and the 
Ryle observations at 2~cm had not yet begun at this time.  The outburst 
centered near day 200$\pm$5 (hereafter ``B'') in the 3.6~cm light curve (Fig. 
1{\it b}) has a maximum amplitude of $\sim$170 mJy; it can be clearly 
identified with the 200 mJy peak on day 190$\pm$5 at 2~cm, and plausibly with 
a small maximum near day 205$\pm$5 at 6~cm and with the broad peak between 
days 220--270 at 20~cm.  In both of these cases, the amplitude of the 
variations appear to decrease with increasing wavelength, and the onset of the 
flares at longer wavelengths seems to be somewhat delayed with respect to the 
shorter wavelengths.  Although the sparse data coverage precludes a detailed 
temporal analysis, the flares appear to rise and decay on a timescale of 
1--2 months.  M81 has been monitored only sparsely at the later phases of the 
program; nonetheless, several additional maxima are evident in the data 
sets, especially at the two shortest wavelengths, and these might correspond to
outbursts similar to ``A'' and ``B.''

In addition to the relatively long-timescale, large-amplitude outbursts, 
several episodes of rapid variability were imprinted in the densely sampled 
portion of the 3.6~cm light curve during the first two months (Fig. 1{\it a}).
The source brightness flickers by 30\%--60\% on a timescale of a day or less, 
which implies that the emission originates from regions with dimensions \lax 
0.001 pc.  Note that the amplitude of the short-term variability is 
significantly larger than the measurement uncertainty, even after allowing for 
pointing errors induced by strong winds (see \S\ 2.1).  Crane \etal (1976) 
previously discussed rapid variability of a similar, perhaps somewhat less 
extreme, nature in M81 at this wavelength.  They found the nucleus to vary by 
about 40\% in the course of a week.  Corresponding variations at longer 
wavelengths are less apparent; at 6~cm, rapid fluctuations occur at most at 
a level of $\sim$10\%, and none are significant at 20~cm.

The wavelength dependent variations in flux density naturally lead to 
strong spectral variations during the outbursts.  Figure 2 displays the 
time variation of the spectral index, $\alpha$, defined such 
that $S_{\nu}\,\propto\, \nu^{\alpha}$, where $S_{\nu}$ is the flux 
density at frequency $\nu$.  As has been well established (de~Bruyn \etal 
1976; Bartel \etal 1982; Reuter \& Lesch 1996), the M81 radio core during 
quiescence has a flat to slightly inverted spectrum; we measure 
$\alpha\,\approx$ 0 to $+$0.3 from 2 to 20~cm, consistent with previous 
determinations.

\begin{figure}
%\plotone{figs/fig2_v5_portrait.ps}
\plotone{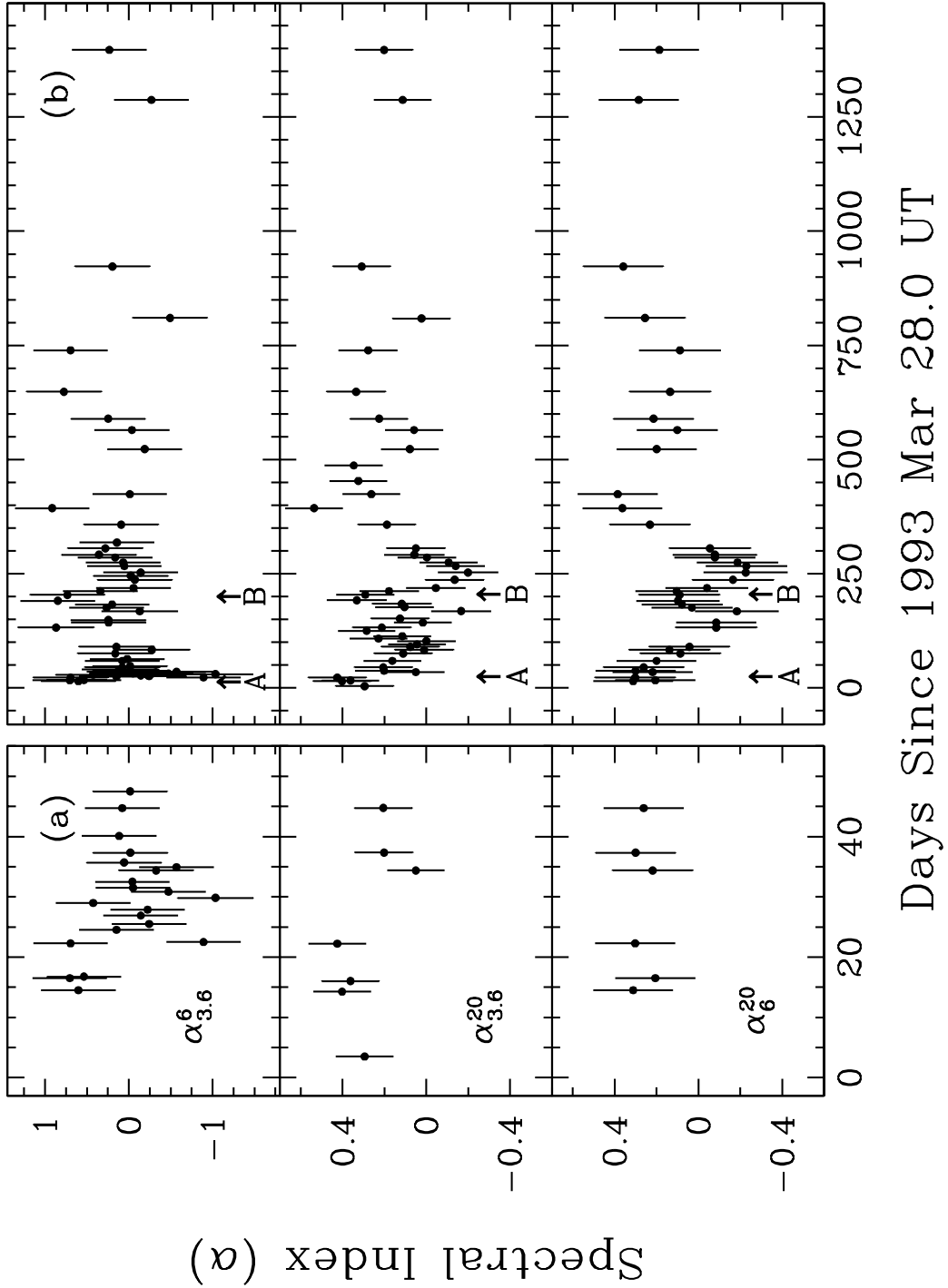}
\caption{
Variation in the radio spectral index $\alpha$, where $S_{\nu}\,\propto\,
\nu^{\alpha}$.  Three indices are shown: $\alpha^6_{3.6}$ = $\alpha$ between
6 and 3.6~cm; $\alpha^{20}_{3.6}$ = $\alpha$ between 20 and 3.6~cm; and
$\alpha^{20}_6$ = $\alpha$ between 20 and 6~cm.  The data for the first
55 days ({\it a}) are shown separately from the full data set ({\it b}).
Outbursts A and B are marked with arrows.
}
\end{figure}

We are convinced that the variations in flux density with time of the nucleus 
are real and intrinsic to the source.  The dotted lines in Figure 1{\it b} 
trace the observed light curves of SN 1993J as derived from the same images 
from which we measured the nucleus.  The supernova light curves exhibit no 
unusual behavior and are well represented by the model of the emission 
discussed in Van Dyk et al. (1994).  The flux and phase calibrations are 
therefore reliable, and we cannot attribute the changes in the source to 
unsuspected variability in the calibrators.  
The effect of sidelobe contamination by the supernova should not be 
significant, as SN 1993J is no stronger than the nucleus, and, moreover, 
the sidelobes are greatly reduced in the cleaned maps. 
In the case of the VLA data, we have properly corrected for primary-beam 
attenuation, we have taken into account the effects of bandwidth smearing, 
and we have carefully considered all known sources of errors affecting the 
flux density measurements.  The only source of uncertainty we did not include 
formally into our error budget is that potentially arising from systematic 
pointing errors induced by wind and solar heating (see \S\ 2.1). However, we 
believe that the main outbursts cannot be attributed to systematic errors 
for the following reasons.  First, the observed variations are much larger 
than even the most pessimistic error estimate.  Second, the main outbursts are 
clearly well resolved in time and so do not arise from any single, errant data 
point.  And finally, one of the outbursts is seen both in the VLA {\it and} in 
the Ryle data sets, which implies that it cannot be attributed to artifacts 
of primary-beam correction of the VLA data.  We further note that the 
variations cannot be a configuration-related effect, since the flux densities, 
particularly at 3.6~cm, do not achieve the same level of variation for 
observations made in the same array configuration but separated in time.  

\section{The Radio Variability and Its Implications}

Previous radio work by Crane \etal (1976) and de~Bruyn \etal (1976) showed 
that the nucleus of M81 undergoes gradual flux variations over several years, 
accompanied by erratic changes on much shorter timescales.  Moreover, these 
authors recorded the onset of a flare in 1974 October; the flux density at 
8085 MHz increased by $\sim$40\% over one week.  The limited time coverage did 
not permit the entire event to be monitored, however, although simultaneous 
observations at 2695 MHz suggested that the flux enhancement might have 
occurred first at the higher frequency. Similarly, Kaufman \etal (1996) 
suggested that a modest flare at 6~cm possibly occurred during 1981 August.  
The observations presented in this paper establish conclusively that the 
nucleus of M81 is strongly variable at centimeter wavelengths.  Our time 
coverage enabled us to identify rapid variability on timescales as short as 
one day or less at 3.6~cm, and more spectacularly, two distinctive outbursts 
which could be traced in more than one wavelength, as well as several others 
seen in at least one of the shorter wavelengths monitored.

Several properties of the best-defined outburst (``B''), namely the steep rise 
and decline of the light curve, and the frequency dependence of the burst 
onset and the burst maximum, suggest that the standard adiabatic-expansion 
model for variable radio sources (Pauliny-Toth \& Kellermann 1966; van der 
Laan 1966) may be applicable.  This model idealizes the radio flux variability 
as arising from an instantaneous injection of a cloud of relativistic 
electrons which is uniform, spherical, and expanding at a constant velocity.  
The flux density increases with source radius (or time) while the source 
remains optically thick, and it decreases during the optically thin phase of 
the expansion.  Both the maximum flux density and the frequency at which it 
occurs decrease with time.  The observed characteristics of outburst ``B'', 
however, do not agree in detail with the predictions of this simple model, as 
is also the case for other variable radio sources (see discussion in 
Kellermann \& Owen 1988).  In particular, the observed profile of the burst at 
2 and 3.6~cm is much shallower than predicted; it follows roughly a linear 
rise and a linear decline, whereas an adiabatically expanding source should 
brighten as $t^3$ before the maximum and, for electrons having a power-law 
energy distribution with a slope of --2.5, decline thereafter as $t^{-5}$.  
Moreover, the 3.6 and 6~cm peaks occur much earlier than expected relative to 
the 2~cm peak, and the relative strengths of the peaks do not follow the 
predicted scaling relations.  These inconsistencies no doubt reflect the 
oversimplification of the standard model.  Realistic modeling of the M81 
source, which is beyond the scope of this work, most likely will need to 
incorporate more complex forms of the particle injection rate (see, e.g., 
Peterson \& Dent 1973), as well as departures from spherical symmetry, since 
the radio source is known to have an elongated geometry, plausibly interpreted 
as a core-jet structure (Bartel \etal 1982; Bartel, Bietenholz, \& Rupen 1995; 
Bietenholz \etal 1996; Ebbers \etal 1998).  The nuclear jet model of Falcke 
(1996), for instance, can serve as a useful starting point.

It is instructive to consider whether the radio variability in M81 is 
associated with any other visible signs of transient activity, at either 
radio or other wavelengths, during the monitoring period.  By analogy with 
other accretion-powered systems, such as Galactic superluminal sources (see, 
e.g., Harmon \etal 1997), one might expect the radio outbursts in the M81 
nucleus to be accompanied by detectable changes in its radio structure and to 
be preceded by X-ray flares.  The nucleus has been intensively imaged at 
milli-arcsec resolution in concert with VLBI studies of SN 1993J.  Bietenholz, 
Bartel, \& Rupen (1998) and Ebbers \etal (1998) do, in fact, report structural 
changes in the jet component of the nucleus at 3.6~cm on timescales of 
weeks, although the limited VLBI data do not permit a direct comparison with 
our VLA light curves.  Several measurements of the nuclear flux in the hard 
X-ray band (2--10 keV) were taken by {\it ASCA} between 1993 May and 1995 
April (Ishisaki \etal 1996), but again, because of the limited temporal 
coverage, these data cannot be used to draw any meaningful comparisons with 
the radio light curve.  

We mention, however, a possible connection between the radio outbursts and an 
optical flare that was caught during the same monitoring period.  After 15 
years of relative constancy (Ho \etal 1996), the broad H\al\ emission line of
the nucleus of M81 brightened by $\sim$40\% and developed a pronounced 
double-peaked line profile (Bower \etal 1996) reminiscent of those seen in a 
minority of AGNs (Eracleous \& Halpern 1994 and references therein).  Neither 
the exact date of its onset nor its time evolution is known, except that it 
occurred between 1993 April 14 (when it was last observed by Ho et al.) and 
1995 March 22 (the date of Bower et al.'s observations), within the period of 
the radio monitoring.  The physical origin of double-peaked broad emission 
lines in AGNs is not yet fully understood (see Halpern \etal 1996 for a 
discussion), and the detection of possibly related variable radio emission 
does not offer a clear discriminant between the main competing models.  
Nevertheless, the association of the sudden appearance of the double-peaked 
line with another transient event, namely the radio outbursts, hints that the 
two events could have a common origin.  Both phenomena, for example, at least 
in this case, may originate from a sudden increase in the accretion rate.

Finally, we note that radio outbursts may be a generic property of 
low-luminosity AGNs, especially those classified as LINERs.  Although no other 
nearby LINER has had its variability properties scrutinized to the same degree
as M81, radio flares have been noticed in at least three other famous LINER 
nuclei.  NGC 1052 is known to have experienced two outbursts at millimeter
wavelengths (Heeschen \& Puschell 1983) and another at longer wavelengths (Slee 
\etal 1994).  A single outburst at centimeter wavelengths has been reported 
for the nucleus of M87 (Morabito, Preston, \& Jauncey 1988).  And Wrobel \& 
Heeschen (1991) remark that NGC 4278 exhibited pronounced variability at 6~cm 
over the course of 1--2 years.  In this regard, it is appropriate to mention 
that even the extremely low-power radio source in the center of the Milky Way, 
Sgr A$^*$, showcases outbursts at high radio frequencies (Zhao \etal 1992; 
Wright \& Backer 1993).  The radio variability characteristics of these weak 
nuclei closely mimic those of far more powerful radio cores traditionally 
studied in quasars and radio galaxies, and they furnish additional evidence 
that the AGN phenomenon spans an enormous range in luminosity.
 
\section{Summary}

We analyzed the radio light curves of the low-luminosity active nucleus 
in the nearby spiral galaxy M81 taken at 3.6, 6, and 20~cm over a four-year 
period between 1993 and 1997, as well as a 2~cm light curve covering a more 
limited span between 1993 and 1994.  Two types of variability are seen: 
rapid (\lax 1 day), small-amplitude (10\%--60\%) flux density changes are 
evident at 3.6 and 6~cm, and at least one, and possibly three or four longer 
timescale (months), outbursts of greater amplitude (30\%--100\%).  The best 
observed of the outbursts can be traced in three bands.  The maximum flux 
density decreases systematically with decreasing frequency, and the time at 
which the maximum occurs is shifted toward later times at lower frequencies.  
These characteristics qualitatively agree with the predictions of the 
adiabatic-expansion model for variable radio sources, although certain 
discrepancies between the observations and the model predictions suggest that 
the model needs to be refined.  The radio outbursts may be related to an 
optical flare during which the broad H\al\ emission line developed a 
double-peaked structure.  Although the exact relationship between the two 
events is unclear, both phenomena may stem from a sudden increase in the 
accretion rate.

\acknowledgments
During the course of this work, L.~C.~H. was supported by a 
postdoctoral fellowship from the Harvard-Smithsonian Center for Astrophysics, 
by NASA grant NAG 5-3556, and by NASA grants GO-06837.01-95A and 
AR-07527.02-96A from the Space Telescope Science Institute (operated by AURA, 
Inc., under NASA contract NAS5-26555).  K.~W.~W. wishes to acknowledge the 
Office of Naval Research for the 6.1 funding which supports his work.  We 
thank Norbert Bartel, Michael Eracleous, Heino Falcke, and the referee for 
helpful comments, and Rick Perley and Michael Rupen for advice concerning 
pointing errors of the VLA.

%\clearpage
\appendix
\section{The Data}

For the sake of completeness, we list in Tables 1--4 the flux densities 
of the nucleus of M81.  The uncertainties in the flux densities were 
calculated as described in \S\ 2.1.  These are the data plotted in Figure 1.

\begin{figure}
\plotone{table1_v4.epsi}
\end{figure}
 
\begin{figure}
\plotone{table2_v4.epsi}
\end{figure}
 
\begin{figure}
\plotone{table3_v4.epsi}
\end{figure}
 
\begin{figure}
\plotone{table4_v4.epsi}
\end{figure}

%REFERENCES
\clearpage

\centerline{\bf{References}}
\medskip

\refindent
Baars, J.~W.~M., Genzel, R., Pauliny-Toth, I.~I.~K., \& Witzel, A., 1977, \aa,
61, 99

\refindent
Bartel, N., \etal 1982, \apj, 262, 556
 
\refindent
Bartel, N., Bietenholz, M.~F., \& Rupen, M.~P. 1995, in Proc. Natl. Acad. 
Sci., 92, 11374

\refindent
Bietenholz, M.~F., \etal 1996, \apj, 457, 604 

\refindent
Bietenholz, M.~F., Bartel, N., \& Rupen, N.~P. 1998, in IAU Colloq. 164,
Radio Emission from Galactic and Extragalactic Compact Sources, ed. A. Zensus,
G. Taylor, \& J. Wrobel (San Francisco: ASP), 201

\refindent
Bower, G.~A., Wilson, A.~S., Heckman, T.~M., \& Richstone, D.~O. 1996, \aj,
111, 1901

\refindent
Crane, P.~C., Giuffrida, B., \& Carlson, J.~B. 1976, \apj, 203, L113
 
\refindent
de~Bruyn, A.~G., Crane, P.~C., Price, R.~M., \& Carlson, J. 1976, \aa, 46, 243
 
\refindent
Ebbers, A., Bartel, N., Bietenholz, M.~F., Rupen, N.~P., \& Beasley, A.~J. 
1998, in IAU Colloq. 164, Radio Emission from Galactic and Extragalactic 
Compact Sources, ed. A. Zensus, G. Taylor, \& J. Wrobel (San Francisco: ASP), 
203

\refindent
Eracleous, M., \& Halpern, J.~P. 1994, \apjs, 90, 1

\refindent
Falcke, H. 1996, \apj, 464, L67

\refindent
Filippenko, A.~V., \& Sargent, W.~L.~W. 1988, \apj, 324, 134 
 
\refindent
Freedman, W.~L., \etal 1994, \apj, 427, 628

\refindent
Halpern, J.~P., Eracleous, M., Filippenko, A.~V., \& Chen, K. 1996, \apj, 464,
704

\refindent
Harmon, B.~A., Deal, K.~J., Paciesas, W.~S., Zhang, S.~N., Robinson, C.~R.,
Gerard, E., Rodr\'\i guez, L.~F., \& Mirabel, I.~F. 1997, \apj, 477, L85

\refindent
Heckman, T.~M. 1980, \aa, 87, 152

\refindent
Heeschen, D.~S., \& Puschell, J.~J. 1983, \apj, 267, L11

\refindent
Ho, L.~C. 1999, in  The AGN-Galaxy Connection, ed. H.~R. Schmitt, L.~C. Ho, \&
A.~L. Kinney (Advances in Space Research), in press (astro-ph/9807273)

\refindent
Ho, L.~C., Filippenko, A.~V., \& Sargent, W.~L.~W. 1996, \apj, 462, 183

\refindent
Ho, L.~C., Filippenko, A.~V., \& Sargent, W.~L.~W. 1997, \apjs, 112, 315

\refindent
Ishisaki, Y., \etal 1996, PASJ, 48, 237

\refindent
Jones, M.~E. 1991, in IAU Colloq. 131, Radio Interferometry: Theory, 
Techniques, and Applications, ed. T.~J. Cornwell \& R.~A. Perley 
(San Francisco: ASP), 295

\refindent
Kaufman, M., Bash, F.~N., Crane, P.~C., \& Jacoby, G.~H. 1996, \aj, 112, 1021

\refindent
Keel, W.~C. 1989, \aj, 98, 195
 
\refindent
Kellermann, K.~I., \& Owen, F.~N. 1988, in Galactic and Extragalactic Radio 
Astronomy, ed. G.~L. Verschuur \& K.~I. Kellermann (New York: Springer-Verlag),
563

\refindent
Morabito, D.~D., Preston, R.~A., \& Jauncey, D.~L. 1988, \aj, 95, 1037

\refindent
Morris, D. 1991, VLA Test Memorandum No. 182 (National Radio Astronomy 
Observatory)

\refindent
Napier, P.~J., \& Rots, A.~H. 1982, VLA Test Memorandum No. 134
(National Radio Astronomy Observatory)

\refindent
Pauliny-Toth, I.~I.~K., \& Kellermann, K.~I. 1966, \apj, 146, 634

\refindent
Peimbert, M., \& Torres-Peimbert, S. 1981, ApJ, 245, 845
 
\refindent
Peterson, F.~W., \& Dent, W.~A. 1973, \apj, 186, 421

\refindent
Pooley, G.~G., \& Green, D.~A. 1993, \mnras, 264, L17

\refindent
Reuter, H.-P., \& Lesch, H. 1996, \aa, 310, L5

\refindent
Shuder, J.~M., \& Osterbrock, D.~E. 1981, \apj, 250, 55

\refindent
Slee, O.~B., Sadler, E.~M., Reynolds, J.~E., \& Ekers, R.~D. 1994, \mnras, 269,
928

\refindent
van der Laan, H. 1966, \nat, 211, 1131

\refindent
Van Dyk, S.~D., Weiler, K.~W., Sramek, R.~A., Rupen, M.~P., \& Panagia,
N. 1994, \apj, 432, L115

\refindent
Weiler, K.~W., Sramek, R.~A., Panagia, N., van der Hulst, J.~M., \& 
Salvati, M. 1986, \apj, 301, 790

\refindent
Wright, M.~C.~H., \& Backer, D.~C. 1993, \apj, 417, 560
 
\refindent
Wrobel, J.~M., \& Heeschen, D.~S. 1991, \aj, 101, 148

\refindent
Zhao, J.-H., Goss, W.~M., Lo, K.~Y., \& Ekers, R.~D. 1992, in Relationships
Between Active Galactic Nuclei and Starburst Galaxies, ed. A.~V. Filippenko
(San Francisco: ASP), 295

\end{document}